\documentstyle[emulateapj]{article}
\input{psfig.sty}

\catcode`\@=11 
\def\@versim#1#2{\vcenter{\offinterlineskip
        \ialign{$\m@th#1\hfil##\hfil$\crcr#2\crcr\sim\crcr } }}
\newcommand{\beq}{\begin{equation}}
\newcommand{\eeq}{\end{equation}}
\def\lsim{\mathrel{\mathpalette\@versim<}}
\def\gsim{\mathrel{\mathpalette\@versim>}}

\newcommand{\kmsec}{\,{\rm km\,s^{-1}}}
\newcommand{\msun}{\,{\rm M_\odot}}
\newcommand{\etal}{{et al.\ }}
\def\spose#1{\hbox to 0pt{#1\hss}}
\newcommand{\lta}{\mathrel{\spose{\lower 3pt\hbox{$\mathchar"218$}}
      \raise 2.0pt\hbox{$\mathchar"13C$}}}
\newcommand{\gta}{\mathrel{\spose{\lower 3pt\hbox{$\mathchar"218$}}
      \raise 2.0pt\hbox{$\mathchar"13E$}}}
\lefthead{Madau \& Quataert}
\righthead{Gravitational recoil and massive black holes}

\makeatletter

\newenvironment{figurehere}
  {\def\@captype{figure}}
  {}
\makeatother

\begin{document}
\submitted{To appear in the ApJL}
\title{The effect of gravitational-wave recoil on the demography of massive 
black holes}
\author{Piero Madau}
\affil{Department of Astronomy \& Astrophysics, University of California, 
Santa Cruz, CA 95064; pmadau@ucolick.org}
\and
\author{Eliot Quataert}
\affil{Department of Astronomy, 601 Campbell Hall, University of California at Berkeley, Berkeley, CA 94720, USA; eliot@astron.berkeley.edu}
               
\begin{abstract}
The coalescence of massive black hole (MBH) binaries following galaxy
mergers is one of the main sources of low-frequency gravitational
radiation.  A higher-order relativistic phenomenon, the recoil as a
result of the non-zero net linear momentum carried away by
gravitational waves, may have interesting consequences for the
demography of MBHs at the centers of galaxies. We study the dynamics
of recoiling MBHs and its observational consequences. The
``gravitational rocket'' may: i) deplete MBHs from late-type spirals,
dwarf galaxies, and stellar clusters; ii) produce off-nuclear quasars,
including unusual radio morphologies during the recoil of a radio-loud
source; and iii) give rise to a population of interstellar and
intergalactic MBHs.

\end{abstract}
\keywords{black hole physics -- cosmology: theory -- galaxies: nuclei -- 
stellar dynamics}

\section{Introduction}
\label{sec:intro}

The first massive black holes likely formed at high redshifts ($z\gsim
10$) at the centers of low-mass ($\sim 10^6{\rm M_\odot}$) dark matter
concentrations.  These black holes evolve into the population of
bright quasars observed at $z \lsim 6$ and eventually grow into the
supermassive remnants that are ubiquitous at the centers of galaxies
in the nearby universe.  In popular cold dark matter (CDM)
cosmogonies, dark matter halos and their associated galaxies undergo
many mergers as mass is assembled from high redshift to the
present. The merging -- driven by dynamical friction against the dark
matter -- of two comparable-mass halo$+$MBH systems will drag in the
satellite hole towards the center of the more massive progenitor,
leading to the formation of a bound MBH binary with separation of
$\sim$ pc. If stellar dynamical and/or gas processes drive the binary
sufficiently close ($\lsim 0.01$ pc), gravitational radiation will
eventually dominate angular momentum and energy losses and cause the
two MBHs to coalesce.  Such catastrophic events are one of the primary
motivations for low-frequency, space-based, gravitational wave
detectors such as the planned {\it Laser Interferometer Space Antenna}
({\it LISA}). For unequal mass pairs, gravitational waves also remove
net {\it linear momentum} from the binary and impart a ``kick''
velocity to the center of mass of the system. The dominant recoil
effect arises from the interference between the mass-quadrupole and
mass-octupole or current-quadrupole contributions (Peres 1962;
Bekenstein 1973).  To date, the outcome of this ``gravitational
rocket'' remains uncertain, as fully general relativistic numerical
computations of radiation reaction effects during the coalescence of
two Kerr holes are not available. For sufficiently asymmetric
configurations, recoil velocities may exceed a few hundred kilometers
per second and lead to a significant displacement of the MBH from the
center of its host galaxy. In the shallow potential wells of
small-mass halos at high redshifts, recoil velocities may be so large
in the late stage of black hole-black hole coalescence to reach
galactic escape velocities (Madau et al. 2004). If it is not ejected
from the host altogether (giving origin to a population of {\it
intergalactic} MBHs), the hole will return to the nucleus via
dynamical friction.

In this {\it Letter} we address the dynamics of recoiling holes in
galaxy cores, discuss the implications of coalescence-induced kicks
for the demography of MBHs, and consider the prospects for directly
detecting the observational signatures of gravitational-wave recoil.
While preparing this work for submission, we learned of an independent
study by Merritt \etal (2004) of the consequences of the gravitational
rocket.

\section{Dynamics of recoiling black holes}

Gravitation radiation recoil is a strong field effect that depends on
the lack of symmetry in the binary system. The lighter hole in a
quasi-circular inspiraling orbit moves faster than the heavier one,
and its gravitational radiation is more ``forward beamed''. This gives
a net momentum ejection in the direction of motion of the lighter
mass, and the binary recoils in the opposite direction (Wiseman
1992). According to quasi-Newtonian calculations (Fitchett 1983), at
the transition from the inspiral to ``plunge'' phase, the center of
mass of a compact binary of total mass $M=m_1+m_2$ (with the
convention $m_1<m_2$) recoils with a velocity \beq v_{\rm
CM}=1480\,\kmsec {f(q)\over f_{\rm max}}\, \left({2GM/c^2\over r_{\rm
isco}}\right)^4, \eeq where the function $f(q\equiv
m_1/m_2)=q^2(1-q)/(1+q)^5$ reaches a maximum value $f_{\rm
max}=0.0179$ for $q=1/2.6$, and $r_{\rm isco}$ is the radius of the
innermost stable circular orbit, which moves inwards for a
comparable-mass system relative to its test-mass limit (e.g. Buonanno
\& Damour 2000). By symmetry, the recoil vanishes for equal-mass
nonrotating holes. Fitchett \& Detweiler (1984) extended Fitchett's
work to perturbation theory and estimated kick velocities $\gta
100\,\kmsec$.  More recent perturbation theory calculations by Favata,
Hughes, \& Holz (2004) find that the recoil velocity can readily reach
$\approx 100-200\,\kmsec$, but is unlikely to exceed $\approx
500\,\kmsec$.  Numerical relativistic calculations of radiation recoil
from highly distorted Schwarzschild holes yields maximum kick
velocities in excess of $400\,\kmsec$ (Brandt \& Anninos
1999). Because of the strong dependence of the rocket effect on the
radius of the final prior-to-plunging orbit, larger kicks are expected
for prograde inspiral into rapidly rotating MBHs.  Moreover, in the
case of Kerr holes, recoil is significant even for holes of equal mass
(Favata et al. 2004) and can be directed out of the plane of the orbit
(Redmount \& Rees 1989).

What is the dynamics of a recoiling MBH in the gravitational potential
of its host galaxy?  As the MBH (with mass $M_{\rm BH}$) recoils from
the nucleus, stars bound to the hole are displaced with it, so that
the total ejected mass is $M_{\rm tot} = M_{\rm BH} + M_{\rm cusp}$,
where the bound stellar cusp has a mass $M_{\rm cusp} \approx M_{\rm
BH}$ for $v_{\rm CM} \approx \sigma$, while $M_{\rm cusp}$ rapidly
decreases for $v_{\rm CM} \gsim \sigma$. As a first approximation let
us assume that the hole + stellar cusp is on a radial orbit in a
spherical potential (e.g., an early-type galaxy or the bulge of a
spiral).  Its orbit is then governed by
\begin{eqnarray} {d^2 {\vec r} \over d t^2} &=&  -{G M(r) \over r^2} 
{\hat r} \nonumber \\ &-& {4 \pi G^2 \ln\Lambda\,\rho M_{\rm tot} \over
v^2} \left({\rm erf}(x) - {2 x \over \sqrt{\pi}} e^{-x^2}\right) {\hat
v},
\label{DF}
\end{eqnarray}
where $x = v/\sqrt{2}\sigma$, $M(r)$ describes the mass profile of the
galaxy, $\rho(r)$ is the density profile of stars with 1D velocity
dispersion $\sigma$, and the second term represents dynamical friction
against the stellar background (e.g. Binney \& Tremaine 1987;
hereafter BT).  We approximate the stellar density as an isothermal
sphere, in which case the dynamical friction time scales with radius
as $t_{\rm DF} \propto r^2$, while the `residence' time for a radial
orbit scales as $t \approx r/v \propto r$.  Thus most of the decay of
the MBH's orbit by friction takes place at small radii in the galactic
nucleus, where it is reasonable to assume that the gravitational
potential is dominated by stars. Stars within the gravitational sphere
of influence of the hole, $R_{\rm BH} \approx GM_{\rm BH}/\sigma^2$,
are bound to it and do not contribute to dynamical friction.  We
therefore truncate the stellar density inside a core radius $\approx
R_{\rm BH}$, so that $\rho(r) = \sigma^2/[2\pi G(r^2 + R_{\rm BH}^2)]$
and $M(r) = 2\sigma^2 r[1 - (R_{\rm BH}/r)\,{\rm arctg}\,(r/R_{\rm
BH})]/G$.  Flattening of the inner stellar density profile is also
produced physically during the decay of the initial MBH binary, when
dynamical friction and three-body interactions transfer energy from
the binary to stars in the nucleus of the galaxy (e.g., Milosavljevic
\& Merritt 2001).

The Coulomb logarithm $\ln \Lambda$ in equation (\ref{DF}) is
approximately given by $\ln[b_{\rm max}/b_{\rm min}]$, where $b_{\rm
max}$ and $b_{\rm min}$ are the maximum and minimum impact parameters
for stars that contribute to dynamical friction (BT); $b_{\rm min}$ is
typically taken to be $\sim R_{\rm BH}$ (BT; Maoz 1993).  In the case
of a uniform medium, $b_{\rm max}$ is comparable to the size of the
system (galaxy), so that $b_{\rm max} \gg b_{\rm min}$. For our
problem, however, the decay of the hole's orbit occurs in the inner
nuclear regions where the stellar density peaks, i.e. at radius $\sim
R_{\rm BH}$, so that $b_{\rm max} \sim R_{\rm BH} \sim b_{\rm min}$
and $\ln \Lambda \sim 1$.  Maoz (1993) has given a more careful
derivation of dynamical friction in an inhomogeneous medium; the
Coulomb logarithm is replaced (see his eq. 4.4) by $\ln \Lambda
\rightarrow \int^{R_{\rm max}}_{R_{\rm BH}} \rho(r) dr/(\rho_0 r)$,
where $R_{\rm max}$ is the size of the stellar system and $\rho_0$ is
the central stellar density.  For our problem this integral is $\sim
1$, confirming the above argument that $\ln \Lambda \approx 1$ is
appropriate.

\begin{figurehere}
\vspace{+0.2cm}
\centerline{
\psfig{file=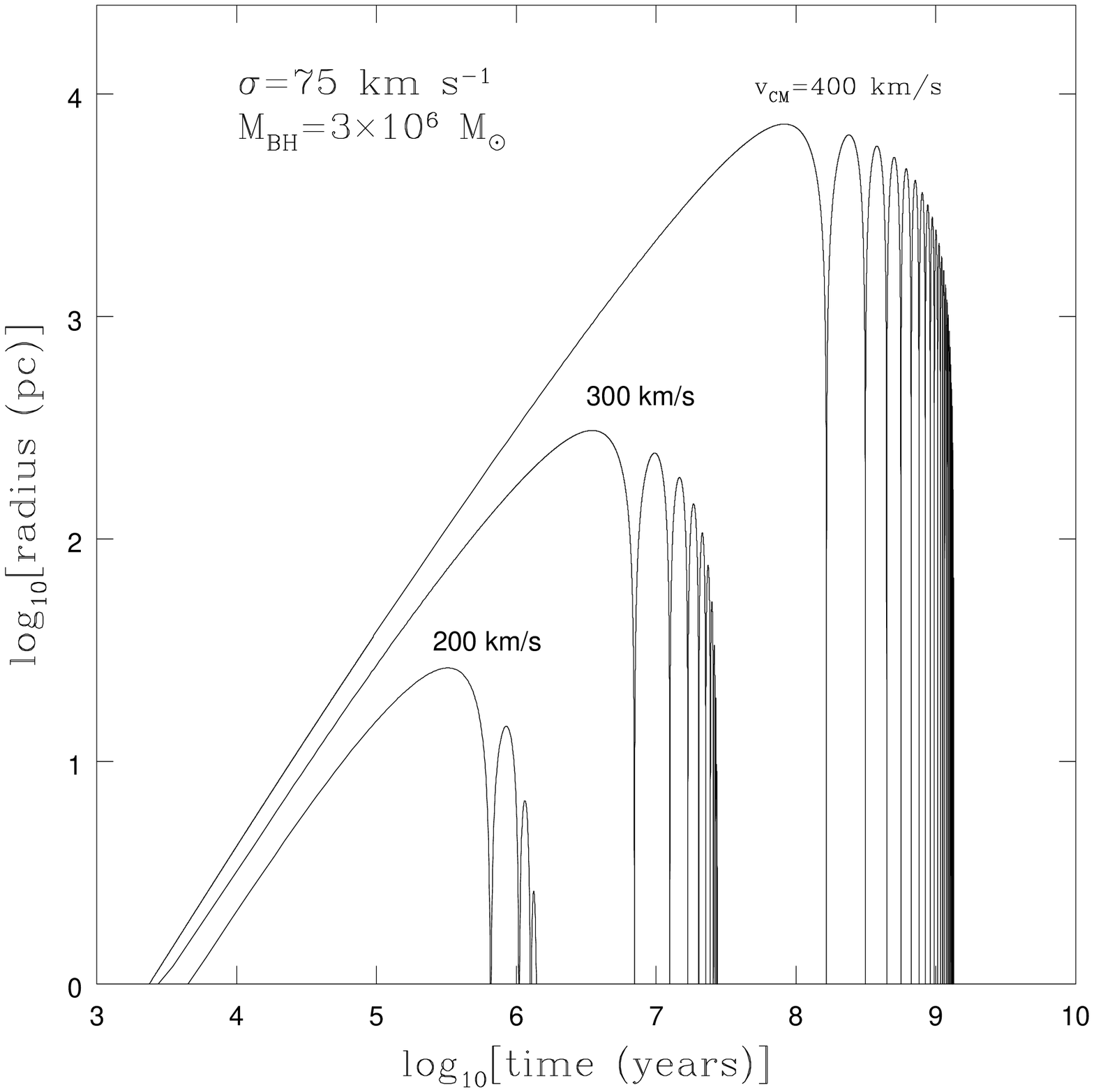,width=3.0in}}
\caption{\footnotesize The response of a $3 \times 10^6 M_\odot$ MBH
to a given recoil velocity $v_{\rm CM}$ in an isothermal potential
with dispersion $\sigma = 75 \kmsec$; radial orbits are assumed.  As
explained in the text, the curves depend primarily on the
dimensionless number $v_{\rm CM}/\sigma$ and so can be scaled to other
parameters.}
\label{fig1} 
\vspace{+0.5cm}
\end{figurehere}

Figure 1 shows the orbit of a $3 \times 10^6 \msun$ MBH moving with
initial recoil velocity of $v_{\rm CM} = 200, 300, {\rm and} \ 400 \,
\kmsec$ in an isothermal sphere of dispersion $\sigma = 75 \kmsec$
(reasonable for a Milky Way-type host) and radius $R_{\rm max}\gg
R_{\rm BH}$. The escape speed from such a sphere is $v_{\rm
esc}(0)=\sqrt{2|\phi|(r=0)}=2\sigma\,\ln^{1/2}(R_{\rm max}/R_{\rm
BH})$.  We fix the hole mass using the $M_{\rm BH}-\sigma$ relation of
Tremaine et al. (2002): $M_{\rm BH} = 1.5 \times 10^8 \sigma_{200}^4 \
M_\odot$, so that $R_{\rm BH} \approx 15\,\sigma_{200}^2$ pc, where
$\sigma_{200}$ is the velocity dispersion in units of $200\,\kmsec$.
As Figure 1 shows, the timescale for the recoiling hole to return to
the center of its host galaxy depends very sensitively on the
magnitude of the kick it receives, ranging from $\sim 10^6$ years for
$v_{\rm CM} =200\,\kmsec$ to more than $10^{9}$ years for $v_{\rm CM}
= 400\,\kmsec$.  Because the friction takes place primarily at small
radii, the decay time is sensitive to the inner stellar density. For
example, if we increase the core radius of the isothermal sphere by a
factor of 2, the decay times increase by a factor of $\approx 3$.
Given uncertainties in the recoil velocity, this uncertainty in the
decay time will not significantly modify our conclusions.

The numerical results in Figure 1 can be understood analytically as
follows. In the absence of dynamical friction, energy is conserved,
and an ejected MBH will reach its apocenter at \beq R_a \sim R_{\rm
BH} \exp\left[\left(v_{\rm CM}/2\sigma\right)^2\right], \label{max}
\eeq where the exponential dependence arises because the potential is
logarithmic.  During each passage through the galactic nucleus,
however, the hole loses a fraction $f$ of its orbital energy because
of dynamical friction; this fraction can be estimated by comparing the
DF timescale at $\sim R_{\rm BH}$ to the transit time across this
region, $\sim R_{\rm BH}/v_{\rm CM}$, which yields $f \sim
(\sigma/v_{\rm CM})^4$.  It takes $\sim f^{-1}$ orbits through the
nucleus for the MBH to return to the center of the galaxy, and so the
decay time can be roughly estimated by $t_{\rm decay} \sim f^{-1}
R_{\rm a}/v_{\rm CM}$.  Since both $f$ and $R_{\rm a}$ depend
primarily on $v_{\rm CM}/\sigma$, this estimate shows that the decay
time of the hole back to the center is determined primarily by $v_{\rm
CM}/\sigma$ for radial orbits.  Thus, the results in Figure 1 can be
scaled to other parameters, e.g. the decay times in Figure 1 apply
reasonably accurately to $\sigma = 50 \kmsec$, so long as $v_{\rm CM}$
is rescaled to $v_{\rm CM} = 266, 200, {\rm and}\ 133 \kmsec$ (from
top to bottom).\footnote{More accurately, at fixed $v_{\rm
CM}/\sigma$, there remains a weak dependence on $\sigma$ via $t_{\rm
decay} \propto R_{\rm a}/v_{\rm CM} \propto R_{\rm BH}/v_{\rm CM}
\propto \sigma$.}


The above analysis assumes that the MBH remains on a purely radial
orbit.  It is, however, likely that the recoiling hole will gain
angular momentum upon leaving the nucleus (e.g. if the galactic
potential is triaxial): this will modify the timescale for the orbital
decay.  Detailed estimates of the decay time in this case are
uncertain. We can roughly bracket the effect by considering the MBH on
a circular orbit with an energy $(1/2) M_{\rm BH} v_{\rm CM}^2$; for
an isothermal potential, the radius of such an orbit is $ \approx
R_{a}/\sqrt{e}$, where $R_{a}$ is the apocenter radius for a
radial orbit (shown in Figure 1 and estimated in eq. [\ref{max}]). The
decay time is then given by the standard dynamical friction timescale
(BT)
\beq
t_{\rm DF} \approx {10^{10} {\rm
yrs} \over \ln \Lambda} \left(R_{a} \over 1.5 \ {\rm kpc}\right)^2
\left(\sigma \over 75 \kmsec\right)^{-3}
\label{decay}, 
\eeq 
where we have used the $M_{\rm BH}-\sigma$ relation to eliminate the
hole mass in favor of $\sigma$.  Note that the decay time is now
determined by dynamical friction at large radii, rather than small
radii, and so $\ln \Lambda \sim 10$ is probably more appropriate.  For
$v_{\rm CM} \approx \sigma$ the decay time predicted by equation
(\ref{decay}) is comparable to that for radial orbits (a few crossing
times; see Fig. 1) while it can be significantly longer for large 
$v_{\rm CM}/\sigma$ (large apocenter distances).

\section{MBH binaries in hierarchical clustering cosmologies}

In hierarchical clustering scenarios, MBH-MBH binaries are likely
products of galaxy major mergers only. When two
halo$+$MBH systems of (total) mass $M$ and $M_s<M$ merge, the
`satellite' (less massive) progenitor will sink to the center of the
more massive pre-existing system by dynamical friction against the
dark matter; for an isothermal sphere the Chandrasekhar dynamical
friction timescale is \beq t_{\rm DF}=1.65{1+M_s/M \over M_s/M}\,
{1\over H\sqrt{\Delta_{\rm vir}} \ln\Lambda}\Theta
\label{eqtdf}
\eeq (Lacey \& Cole 1993), where $\Delta_{\rm vir}$ is the density
contrast at virialization, $H$ is the Hubble parameter, 
and the term $\Theta$
contains the dependence of this timescale on the orbital
parameters. After including the increase in the orbital decay
timescale due to tidal stripping of the satellite (Colpi \etal 1999),
it is possible to show that satellites will merge with the central
galaxy on timescales shorter than the Hubble time only in the case of
major mergers, $M_s/M\gta 0.3-0.5$. In minor mergers tidal stripping
may leave the satellite MBH wandering in the halo, too far from the
center of the remnant for the formation of a black hole binary.

Major mergers are frequent at early times, so a significant number of
binary MBH systems is expected to form then. It is still unclear
whether halo major mergers necessarily lead to the coalescence of
their MBHs, or whether the binaries ``stall'' before the backreaction
from gravitational waves becomes important (Begelman, Blandford, \&
Rees 1980). A number of plausible mechanisms that may help avoid such
stalling have been suggested in the recent literature (e.g., Gould \&
Rix 2000; Zhao, Haehnelt, \& Rees 2001; Armitage \& Natarajan 2002; Yu
2002; Chatterjee, Hernquist, \& Loeb 2003).

\begin{figurehere}
\vspace{+0.2cm}
\centerline{
\psfig{file=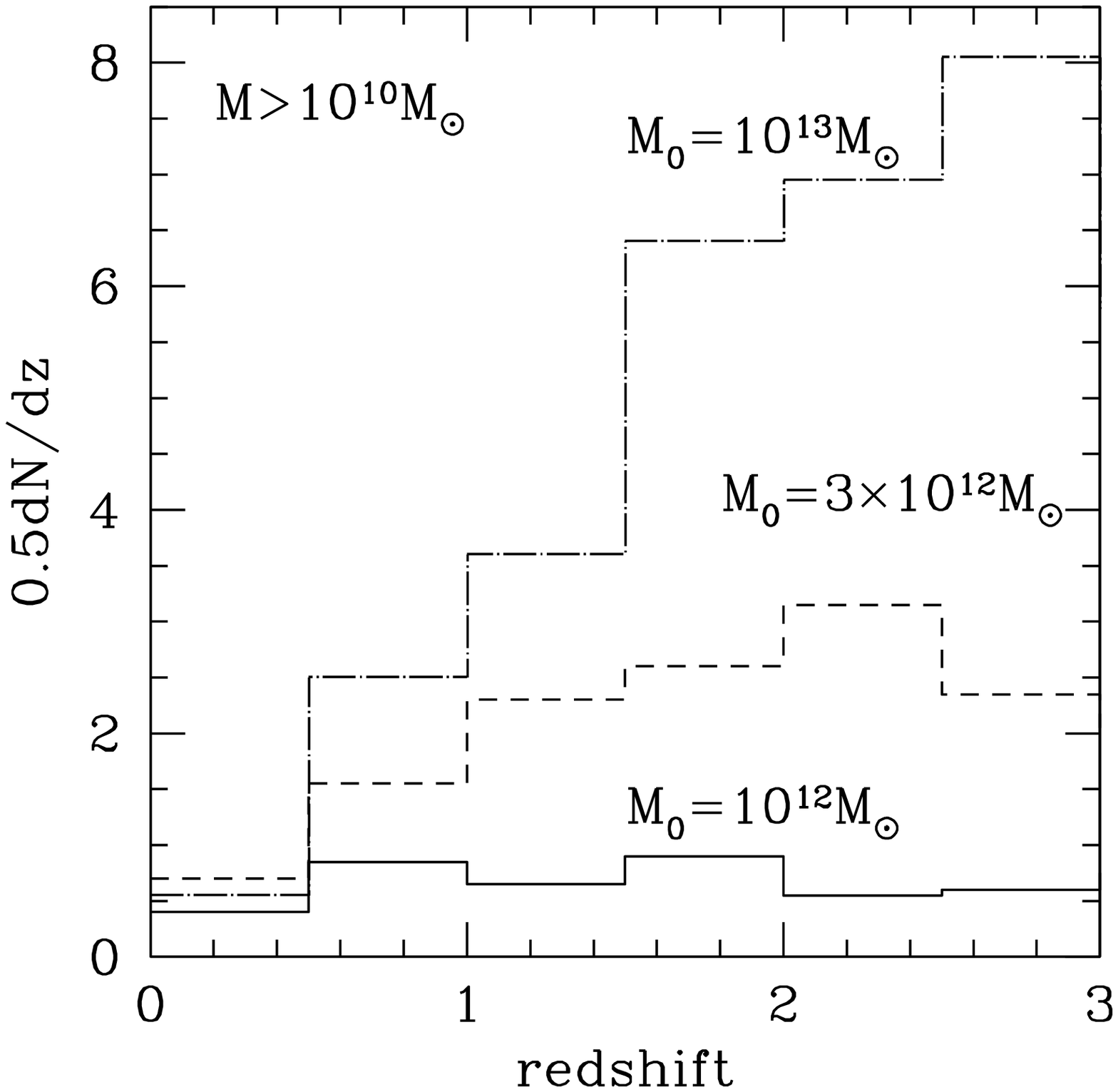,width=3.0in}}
\caption{\footnotesize Mean number of major mergers (with mass ratios $>0.5$) 
experienced 
per unit redshift by halos with masses $>10^{10}\,\msun$. {\it Solid line:}
progenitors of a $M_0=10^{12}\,\msun$ halo at $z=0$. {\it Dashed line:}
same for $M_0=3\times 10^{12}\,\msun$. {\it Dot-dashed line:} same for
$M_0=10^{13}\,\msun$. The rates have been computed assuming the currently 
favoured $\Lambda$CDM world model with $\Omega_M=0.3$, $\Omega_\Lambda=0.7$, 
$h=0.7$, $\Omega_b=0.045$, $\sigma_8=0.93$, and $n=1$.}
\label{fig2} 
\vspace{+0.5cm}
\end{figurehere}

Figure 2 shows the mean number of major mergers per unit redshift bin
experienced by all halos with masses $>10^{10}\,\msun$ that are
progenitors of a $z=0$ parent halo of mass $M_0=10^{12}, 3\times
10^{12}$, and $10^{13}\,\msun$. We have tracked backwards
in time the merger history of parent halos with a Monte Carlo
algorithm based on the extended Press-Schechter formalism (Volonteri,
Haardt, \& Madau 2003).  For the most massive parent halos this
quantity peaks in the redshift range 2-3, the epoch when the observed
space density of optically-selected quasar also reaches a maximum
(Kauffmann \& Haehnelt 2000). Hydrodynamic simulations of major
mergers have shown that a significant fraction of the gas in
interacting galaxies falls to the center of the merged system (Mihos
\& Hernquist 1994): the cold gas may be eventually driven into the
very inner regions, fueling an accretion episode and the growth of the
nuclear MBH.  We will discuss below the possibility that
gravitational-wave recoil may thus give rise to off-nuclear active
galactic nuclei (AGN).

\section{Implications}\label{sec:conclude}

For galaxies with $\sigma \lsim 50-75\,\kmsec$, typical kick
velocities of a few hundred km s$^{-1}$ are sufficient to unbind the
hole or displace it sufficiently from the nucleus that the decay time
due to dynamical friction is comparable to the Hubble time (Fig. 1).
This implies that MBHs with masses $M_{\rm BH} \lsim 10^6 M_\odot$
(using the $M_{\rm BH}-\sigma$ relation) may be comparably rare in
their late-type spiral or dwarf-galaxy hosts. Interestingly, there are
very few observational constraints on the presence of MBHs in such
galaxies, though Filippenko \& Ho (2003) argued for a $\sim 10^5
M_\odot$ MBH for the Seyfert galaxy in the late type (bulgeless)
spiral NGC 4395, and Barth et al. (2004) reached a similar conclusion
for the dwarf Seyfert 1 Galaxy POX 52.  It is important to stress that
even if galaxies do not currently harbor a central MBH, they may have
done so in the past.  Thus the absence of MBHs in shallow potential
wells would not necessarily imply inefficient MBH formation (e.g.,
Haiman, Madau, \& Loeb 1999), but could instead be due to recoil
during MBH coalescence.  Note also that the low-mass black holes that
are preferentially affected by gravitational recoil are also those
that are expected to dominate the {\it LISA} gravitational wave signal
from MBH-MBH coalescence.  If ejection of such MBHs is common, this
may decrease the number of sources detected by {\it LISA}.

Galaxy mergers are a leading mechanism for supplying fuel to MBHs
(e.g., Barnes \& Hernquist 1991), and so a natural implication of
gravitational recoil is the possibility of off-nuclear quasar
activity.\footnote{One complication is that the dominant episode of
accretion onto MBHs during galaxy mergers could happen when the
satellite galaxy is still sinking in towards the nucleus of the more
massive galaxy. In this case the AGN activity could be completed
before the binary actually coalesces.}  Since the lifetime of
merger-driven activity is of order the Salpeter timescale
$t_S=4.5\times 10^7$ yr, off-nuclear AGN are most likely to be found
in relatively small potential wells. This is because, for (say) a
$\sim 10^8 M_\odot$ MBH in a $\sigma = 200\,\kmsec$ galaxy, the decay
timescale due to dynamical friction is $\lsim 10^6$ yr $\ll t_S$ even
for a kick velocity of $v_{\rm CM} = 300\,\kmsec$, and the
displacement from the nucleus is quite small, $\lsim R_{\rm BH}
\approx 10$ pc. By contrast a decay timescale comparable to $\sim t_S$
is plausible for MBHs with $M_{\rm BH} \approx 10^6-10^7 M_\odot$ and
$v_{\rm CM} \approx 200\,\kmsec$.  Since the merger rate for
progenitor halos hosting such MBHs peaks at $z \approx 2-3$ (Fig. 2),
we predict that a significant fraction of {\it moderate- to
high-redshift, low mass} AGN could be off-nuclear.  Most currently
detected AGN at high redshift are the rare $\sim 10^8-10^{10} M_\odot$
holes hosted by massive halos, for which recoil is probably not
important. At early times, however, theoretical models of the
evolution of the quasar population in hierarchical structure formation
scenarios predict a large number of fainter, low-$M_{\rm BH}$ sources
(e.g., Haiman \& Loeb 1998; Haiman et al. 1999; Volonteri \etal 2003).

The effects of gravitational recoil may be particularly prominent in
radio observations of AGN.  Merritt \& Ekers (2002) argued that MBH
coalescence might imprint itself on the morphology of radio galaxies
by changing the spin axis of the MBH and thus the direction of the
radio jet.  Gravitational recoil may have a comparably important
effect by displacing the radio-loud AGN from the nucleus of the
galaxy.  This could manifest itself as a flat spectrum radio core
displaced from the optical nucleus of the galaxy, along with a jet or
lobe still symmetric about the nucleus.  Radio observations also have
the obvious advantages that contamination from the host galaxy is less
important (relative to optical quasars) and VLBI interferometry allows
precise localization of the radio emission.

Gravitational recoil may also be relevant for understanding the origin
of off-nuclear ultraluminous X-ray sources in nearby galaxies (see,
e.g., Colbert \& Mushotzky 1999).  One interpretation of such sources
is that they are intermediate-mass black holes (IMBHs) with $M \sim
10^3 \msun$ accreting near the Eddington limit.  Miller \& Hamilton
(2002) suggest that such intermediate-mass holes could form in
globular clusters by repeated black hole mergers.  The gravitational
rocket will, however, probably prevent substantial growth via mergers
in the shallow potential well of a globular cluster, since even
mergers with mass ratios as small as $\approx 0.1$ can lead to kick
velocities in excess of the escape velocity of the cluster.  It should
be noted that an IMBH could still form via stellar (rather than
compact object) mergers during the core-collapse of young star
clusters (e.g., G\"urkan et al. 2004).


Finally, we draw attention to the fact that the gravitational rocket
does not necessarily limit the ability of MBHs to grow via gas
accretion from rare less massive seeds, such as IMBHs produced by the
collapse of Population III stars at $z\sim 20$ (Madau \& Rees
2001). This is because seed holes that are as rare as (say) the
3.5$\sigma$ peaks of the primordial density field will evolve largely
in isolation, as the merging of two (mini)halos both hosting a black
hole is a rare event at these very early epochs. A significant number
of MBH binary systems may form only later, when the fraction of halos
hosting MBHs is larger. By then the typical host will be further down
the merger hierarchy (more massive) and the effect of radiation recoil
less disruptive (Madau \etal 2004).

\acknowledgements{EQ thanks Chung-Pei Ma for useful conversations and
acknowledges financial support from NASA grant NAG5-12043, NSF grant
AST-0206006, an Alfred P. Sloan fellowship, and the David \& Lucille
Packard Foundation. PM thanks Marta Volonteri for helping with Figure 2 
and acknowledges support from NASA grant NAG5-11513 and NSF grant AST-0205738.}

\end{document}